\documentclass[twocolumn,showpacs,preprintnumbers,amsmath,amssymb]{revtex4}
\usepackage{graphicx}
\usepackage{dcolumn}
\usepackage{bm}

\begin{document}

\title{Roles of deformation and neutron excess on the giant monopole resonance in 
neutron-rich Zr isotopes
}

\author{Kenichi Yoshida}
\affiliation{RIKEN Nishina Center for Accelerator-Based Science, Wako, Saitama 351-0198, Japan
}%

\date{\today}

\begin{abstract}
We investigate the roles of deformation on the giant monopole resonance (GMR), 
particularly the mixing of the giant quadrupole resonance (GQR) 
and the effects of the neutron excess 
in the well-deformed nuclei around $^{110}$Zr 
and in the drip-line nuclei around $^{140}$Zr 
by means of the deformed quasiparticle-random-phase-approximation 
employing the Skyrme and the local pairing energy-density functionals. 
It is found that the isoscalar (IS) GMR has a two-peak structure, 
the lower peak of which is associated 
with the mixing between the ISGMR and the $K^{\pi}=0^{+}$ component of the ISGQR. 
The transition strengths of the lower peak of the ISGMR grows as 
the neutron number increases. 
In the drip-line nuclei, the neutron excitation is dominant over the proton excitation.
We find for an isovector (IV) excitation the GMR has a four-peak structure 
due to the mixing of the IS and IV modes as well as the mixing of the 
$K^{\pi}=0^{+}$ component of the IVGQR. 
Besides the GMR, we find the threshold strength generated by neutrons only.
\end{abstract}

\maketitle

\section{Introduction}
Nuclei far from the stability has been one of the major interests 
in nuclear structure physics because 
exploring the multipole responses in unstable nuclei 
can provide information on collective excitations in asymmetric nuclear systems. 
The surface structure of neutron-rich nuclei 
is quite different from that of stable nuclei, 
because of the presence of the loosely bound neutrons. 
Since collective excitations are sensitive to the surface structure, 
we can expect new kinds of exotic excitation modes  
to appear in neutron-rich nuclei. 

An example of such an exotic excitation mode in unstable nuclei 
is the pygmy dipole resonance (PDR) appearing in the low energy region of 
the giant dipole resonance (GDR)~\cite{paa07}. 
The PDR consists of the isoscalar (IS) and isovector (IV) characters, 
and they are thought to be strongly mixed~\cite{yos09b}. 
Strong mixing of the IS and IV character leads to a possible emergence of the 
neutron- or proton-dominant giant resonance (GR) in unstable nuclei.

GRs are typical collective modes of excitation in nuclei. 
Effects of the nuclear deformation on the GRs has been investigated for long 
both experimentally and theoretically~\cite{har01}. 
The deformation splitting of the GDR has been well established. 
This is due to the oscillation along the major- and minor-axis~\cite{BM2}. 
For the GRs with higher multipolarity, the deformation splitting is less pronounced~\cite{har01}.
Deformation effects on the GMR has been discussed 
soon after the GMR was established~\cite{gar80}.

Collective excitations such as GRs and low-lying vibrational modes 
are described microscopically by the self-consistent Hartree-Fock (HF) 
plus the random-phase approximation (RPA).
There have been many attempts to investigate the new kinds of collective excitations 
in unstable nuclei~\cite{paa07,ben03,paa10}. 
They are however largely restricted to spherical systems, and the collective excitation
modes in deformed neutron-rich nuclei remain mostly unexplored.

Role of deformation on GRs has been studied by means of the deformed quasiparticle-RPA (QRPA) 
employing the Gogny interaction in the light mass region~\cite{per08}. 
Using Skyrme functionals, 
GRs in heavy systems have been investigated by means of the RPA, 
where the separable approximation is employed for the residual interaction~\cite{nes06}. 

The ground-state properties, deformation and superfluidity, in 
neutron-rich Zr isotopes up to the drip line have been studied~\cite{bla05,sto08}, and 
it is predicted that $^{112}$Zr has large deformation, 
and that the drip-line nuclei are also strongly deformed. 
In the present article, we investigate the GMR in 
neutron-rich medium-mass nuclei around $^{110}$Zr and the drip-line nuclei around $^{140}$Zr 
with paying attention to the deformation effects and the mixing of 
the IS and IV characters. 
To this end, we use the newly-developed 
parallelized HFB and deformed-QRPA calculation code~\cite{yos10}. 

The article is organized as follows: 
In Sec.~\ref{method}, the deformed Skyrme-HFB + QRPA method is recapitulated. 
In Sec.~\ref{result}, results of the numerical analysis of the giant resonances 
in deformed neutron-rich Zr isotopes are presented. 
Finally, summary is given in Sec.~\ref{summary}. 

\section{Model}\label{method}
A detailed discussion of the deformed Skyrme-HFB + QRPA can be found in Refs.~\cite{yos08}. 
Therefore, we just briefly recall the outline of the calculation scheme. 

In order to describe simultaneously the nuclear deformation 
and the pairing correlations including the unbound quasiparticle states, 
we solve the HFB equations~\cite{dob84,bul80}
in coordinate space using cylindrical coordinates $\boldsymbol{r}=(\rho,z,\phi)$. 
We assume axial and reflection symmetries.
For the mean-field Hamiltonian $h$, we employ the SkM* interaction~\cite{bar82}. 
The pairing field is treated by using the density-dependent contact
interaction~\cite{cha76}
\begin{equation}
v_{pair}(\boldsymbol{r},\boldsymbol{r}^{\prime})=\dfrac{1-P_{\sigma}}{2}
\left[ t_{0}^{\prime}+\dfrac{t_{3}^{\prime}}{6}\varrho_{0}^{\gamma}(\boldsymbol{r}) \right]
\delta(\boldsymbol{r}-\boldsymbol{r}^{\prime}), \label{pair_int}
\end{equation}
where $\varrho_{0}(\boldsymbol{r})$ denotes the isoscalar density of the ground state 
and $P_{\sigma}$ the spin exchange operator.
Assuming time-reversal symmetry and reflection symmetry with respect to the $x-y$ plane,
we have to solve for positive $\Omega$ and positive $z$ only, 
$\Omega$ being the $z-$component of the angular momentum $j$. 
We use a lattice mesh size $\Delta\rho=\Delta z=0.6$ fm and a box
boundary condition at $\rho_{\mathrm{max}}=14.7$ fm, $z_{\mathrm{max}}=14.4$ fm. 
The differential operators are represented by use of the 11-point formula of finite difference method. 
Since the parity and $\Omega$ are good quantum numbers in the present calculation scheme, 
we have only to diagonalize the HFB Hamiltonian for each $\Omega^{\pi}$ sector. 
The quasiparticle energy is cut off at $E_{\mathrm{qp,cut}}=60$ MeV
and the quasiparticle states up to $\Omega^{\pi}=23/2^{\pm}$ are included.

The pairing strength parameter $t_{0}^{\prime}$ is
determined so as to approximately reproduce the experimental pairing gap of
$^{120}$Sn ($\Delta_{\mathrm{exp}}=1.245$ MeV).
The strength $t_{0}^{\prime}=-240$ MeV fm$^{3}$ for the
mixed-type interaction ($t_{3}^{\prime}=-18.75t_{0}^{\prime}$)~\cite{ben05} 
with $\gamma=1$ leads to the neutron pairing gap
$\langle \Delta_{\nu}\rangle=1.20$ MeV in $^{120}$Sn.

Using the quasiparticle basis obtained
as a self-consistent solution of the HFB equations,
we solve the QRPA equation in the matrix formulation~\cite{row70}
\begin{equation}
\sum_{\gamma \delta}
\begin{pmatrix}
A_{\alpha \beta \gamma \delta} & B_{\alpha \beta \gamma \delta} \\
-B_{\alpha \beta \gamma \delta} & -A_{\alpha \beta \gamma \delta}
\end{pmatrix}
\begin{pmatrix}
X_{\gamma \delta}^{i} \\ Y_{\gamma \delta}^{i}
\end{pmatrix}
=\hbar \omega_{i}
\begin{pmatrix}
X_{\alpha \beta}^{i} \\ Y_{\alpha \beta}^{i}
\end{pmatrix} \label{eq:AB1}.
\end{equation}
The residual interaction in the particle-hole (p-h) channel appearing
in the QRPA matrices $A$ and $B$ is
derived from the Skyrme density functional. 
We neglect 
the Coulomb interaction to reduce the computing time in the QRPA calculation.
We expect that the two-body Coulomb interaction plays only a minor role~\cite{ter05,eba10}. 
We drop the so-called $``{J}^{2}"$ term $C_{t}^{T}$ both in 
the HFB and QRPA calculations. 
The residual interaction in the
particle-particle (p-p) channel is derived from the pairing
functional constructed from the density-dependent contact
interaction (\ref{pair_int}). 
Furthermore, 
we introduce a cut-off for the two-quasiparticle excitation energy as 60 MeV. 

Since the full self-consistency between the static mean-field
calculation and the dynamical calculation is broken by the above
neglected term and the severe cut-off for the two-quasiparticle excitations, 
the spurious states may have finite excitation energies. 
In the present calculation, the spurious states appear at 0.58 MeV and 0.53 MeV 
for the $K^{\pi}=0^{+}$ and $1^{+}$ states, respectively in $^{110}$Zr.
We assume the contamination of the spurious component in GRs to be small 
because the GRs are well apart from the spurious states. 

In the present calculation, 
the number of two-quasiparticle excitations 
for the $K^{\pi}=0^{+}$ excitation in $^{110}$Zr ($^{140}$Zr) is about 36 000 (46 200). 
For the calculation of the matrix elements and diagonalization, 
it takes about 346 (570) CPU hours and 184 (310) CPU hours, respectively. 
 
\begin{table}[t]
\begin{center}
\caption{Ground state properties of $^{100,108,110,112,136,138,140}$Zr obtained by the deformed HFB calculation 
with the SkM* interaction and the mixed-type pairing interaction. Chemical potentials (in MeV), 
deformation parameters, intrinsic quadrupole moments (in fm$^{2}$), 
average pairing gaps (in MeV), root-mean-square radii (in fm) for 
neutrons and protons are listed. }
\label{GS}
\begin{tabular}{cccccccc}
\hline \hline
\noalign{\smallskip}
  & $^{100}$Zr & $^{108}$Zr & $^{110}$Zr & $^{112}$Zr & $^{136}$Zr & $^{138}$Zr & $^{140}$Zr  \\
\noalign{\smallskip}\hline\noalign{\smallskip}
$\lambda_{\nu}$   & $-6.81$ & $-4.57$ & $-4.31$ & $-3.76$ & $-0.51$ & $-0.36$ & $-0.10$  \\
$\lambda_{\pi}$   & $-10.6$ & $-13.6$ & $-14.4$ & $-15.0$ & $-22.3$ & $-22.6$ & $-22.8$ \\
$\beta_{2}^{\nu}$  & 0.38 & 0.37 & 0.37 & 0.39 & 0.28 & 0.29 & 0.31 \\
$\beta_{2}^{\pi}$  & 0.41 & 0.42 & 0.42 & 0.43 & 0.32 & 0.33 & 0.36 \\
$\langle Q \rangle_{\nu}$  & 602 & 791 & 753 & 849 & 947 & 1023 & 1135  \\
$\langle Q \rangle_{\pi}$  & 403 & 429 & 432 & 456 & 355 & 374 & 409  \\
$\langle \Delta \rangle_{\nu}$   & 0.00 & 0.61 & 0.55 & 0.00 & 0.70 & 0.64 & 0.53 \\
$\langle \Delta \rangle_{\pi}$   & 0.32 & 0.00 & 0.00 & 0.00 & 0.36 & 0.36 & 0.36 \\
$\sqrt{\langle r^{2} \rangle_{\nu}}$  & 4.60 & 4.77 & 4.82 & 4.88 & 5.29 & 5.34 & 5.39 \\
$\sqrt{\langle r^{2} \rangle_{\pi}}$  & 4.43 & 4.51 & 4.53 & 4.56 & 4.69 & 4.72 & 4.75 \\
\noalign{\smallskip}
\hline \hline
\end{tabular}
\end{center}
\end{table}

\section{Results and discussion}\label{result}
We summarize in Table~\ref{GS} the ground state properties.
The neutron-rich Zr isotopes under investigation are prolately deformed. 
This is consistent with the results calculated 
using the Skyrme SLy4 interaction~\cite{bla05}, 
and we also find that $^{112}$Zr has largest deformation. 
Both neutrons and protons are unpaired in this nucleus. 
This calculation indicates that $^{140}$Zr is located very close to the drip line; 
the chemical potential is only $-0.1$ MeV. 
The neutron skin well develops in this nucleus; the difference of the radii of 
neutrons and protons is 0.64 fm.

\begin{figure*}[t]
\begin{center}
\includegraphics[scale=1.0]{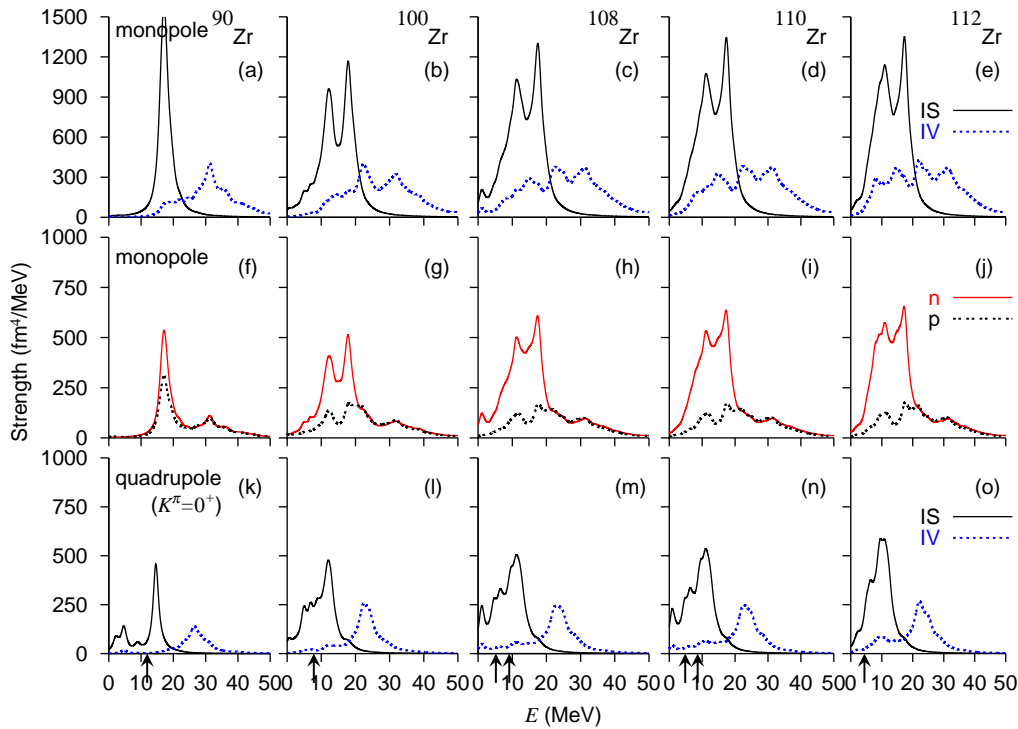}
\caption{(Color online) 
Isoscalar (IS) and isovector (IV) monopole and quadrupole transition strengths 
in $^{90,100,108,110,112}$Zr. 
The neutron and proton transition strengths are also shown for the monopole excitation. 
The arrows indicate the one-quasineutron continuum threshold $E_{\mathrm{th},1n}=|\lambda|+\min E_{\alpha}$ 
and the two-quasineutron continuum threshold $E_{\mathrm{th},2n}=2|\lambda|$. 
Shown is only the one-neutron continuum threshold for $^{90,100,112}$Zr, the normalfluid systems.
}
\label{response1}
\end{center}
\end{figure*}

\begin{figure}[t]
\begin{center}
\includegraphics[scale=1.0]{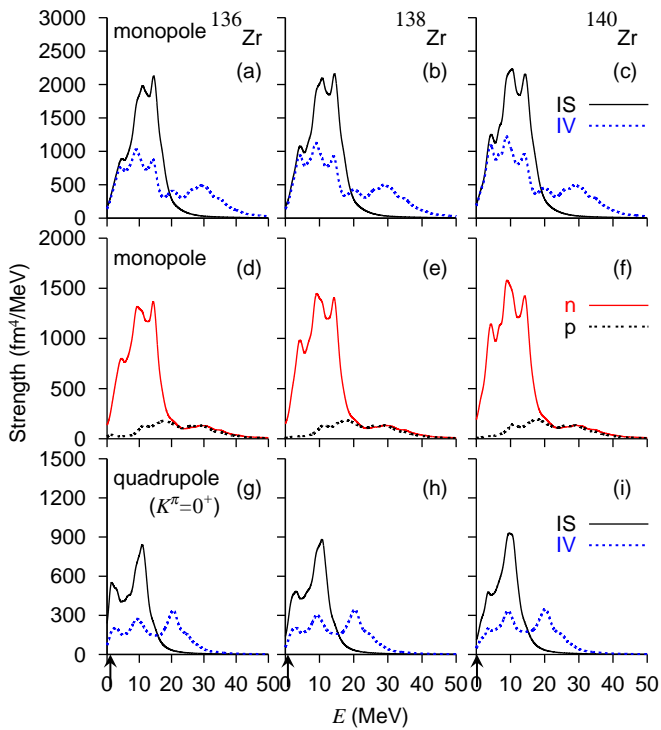}
\caption{(Color online) 
Same as Fig.~\ref{response1} but in $^{136,138,140}$Zr.
}
\label{response2}
\end{center}
\end{figure}

Figures~\ref{response1} and \ref{response2} show the isoscalar (IS) and isovector (IV) 
monopole and quadrupole transition-strength distributions 
in $^{90,100,108,110,112}$Zr and in $^{136,138,140}$Zr. 
Here we define these operators as
\begin{align}
\hat{F}^{\tau=0}_{\lambda=0} &=
\sum_{\tau_{z}=1,-1} \int d\boldsymbol{r} r^{2}\hat{\psi}_{\tau_{z}}^{\dagger}(\boldsymbol{r})
\hat{\psi}_{\tau_{z}}(\boldsymbol{r}), \\
\hat{F}^{\tau=1}_{\lambda=0} &=
\sum_{\tau_{z}=1,-1} \int d\boldsymbol{r} \tau_{z}r^{2}\hat{\psi}_{\tau_{z}}^{\dagger}(\boldsymbol{r})
\hat{\psi}_{\tau_{z}}(\boldsymbol{r}), \\
\hat{F}^{\tau=0}_{\lambda=2,K} &=
\sum_{\tau_{z}=1,-1} \int d\boldsymbol{r} r^{2}Y_{\lambda K}(\hat{r})
\hat{\psi}_{\tau_{z}}^{\dagger}(\boldsymbol{r})
\hat{\psi}_{\tau_{z}}(\boldsymbol{r}), \\
\hat{F}^{\tau=1}_{\lambda=2,K} &=
\sum_{\tau_{z}=1,-1} \int d\boldsymbol{r} \tau_{z}r^{2}Y_{\lambda K}(\hat{r})
\hat{\psi}_{\tau_{z}}^{\dagger}(\boldsymbol{r})
\hat{\psi}_{\tau_{z}}(\boldsymbol{r}).
\end{align} 
A nucleon creation operator $\hat{\psi}^{\dagger}(\boldsymbol{r})$ at the position $\boldsymbol{r}$ is  
written in terms of the quasiparticle wavefunctions.

The strength function is calculated as
\begin{equation}
S_{\lambda}^{\tau}(E)
=\sum_{i}\sum_{K} \dfrac{\Gamma/2}{\pi}\dfrac{|\langle i|\hat{F}_{\lambda K}^{\tau}|0\rangle|^{2}}
{(E-\hbar \omega_{i})^{2}+\Gamma^{2}/4}.
\end{equation} 
The smearing width $\Gamma$ is used to remove the 
spurious oscillation in the strength functions 
due to the box boundary condition, 
and it is set to $\Gamma=2$ MeV in the case of the box size of 15 fm~\cite{ina09}.
 
The ISGMR has a two-peak structure in the deformed nuclei. 
The higher-energy peak of the IS monopole excitation seen around 20 MeV 
is identified as a primal ISGMR 
because the ISGMR in $^{90}$Zr is located at around 18 MeV, 
where we have only one peak.

To understand the origin of the lower-energy peak around 10 MeV, 
we show in Figs.~\ref{response1}(k)-\ref{response1}(o) 
the $K^{\pi}=0^{+}$ component of the quadrupole transition strength. 
At the same energy region where the lower-energy peak of the ISGMR appears, 
we can see a peak structure. 
It is noted here that the neutron-rich Zr isotopes under investigation 
are prolately deformed, so the giant quadrupole resonance (GQR) splits 
into three resonances of the $K^{\pi}=0^{+}, 1^{+}$, and $2^{+}$ components, 
and the resonance peak of the $K^{\pi}=0^{+}$ component lowers in energy. 
 
\begin{figure}[t]
\begin{center}
\includegraphics[scale=0.6]{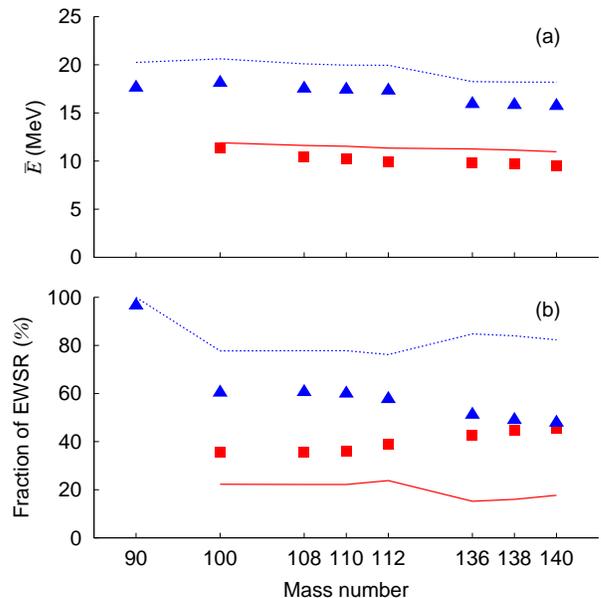}
\caption{(Color online) 
(a) Mean energy of the ISGMR in Zr isotopes. 
The energy of the upper and lower peaks are denoted by triangle and square, respectively. 
(b) Fraction of the EWSR value for the upper and lower peaks. 
The mean energy of the upper and lower peaks 
is calculated as $m_{1}/m_{0}$ for the energy interval of $5$ MeV $<E<E_{c}$ and 
$E_{c} <E<25$ MeV, and the fraction is calculated for each interval. 
The lines represent the results obtained in the scaling model~\cite{nis85}.}
\label{energy}
\end{center}
\end{figure}

We are going to discuss in detail the deformation splitting of the ISGMR. 
Figure~\ref{energy}(a) shows the mean energy of the lower peak and the upper peak 
of the ISGMR. 
The mean energy is evaluated as $m_{1}/m_{0}$, 
the ratio of the energy-weighted sum to the non energy-weighted sum. 
The energy interval is set to $5$ MeV$ < \hbar \omega_{i}< E_{c}$ 
and $E_{c} < \hbar \omega_{i}<25$ MeV for 
the lower and upper peaks. 
The value $E_{c}$ is set to 15 MeV and 14 MeV for $^{100}$Zr and $^{108,110,112}$Zr, respectively. 
We change $E_{c}$ associated with the mass number because 
the resonance energy decreases as the mass number increases. 
The values obtained in the scaling model~\cite{nis85} are also shown in Fig.~\ref{energy}. 

The results of the present calculation are not very different from those 
obtained in the scaling model, with the exception that 
the scaling model overestimates the energy of the compressible mode~\cite{nis85}. 
As the system is deformed from $^{90}$Zr to $^{100}$Zr, the excitation energy 
of the upper peak is upward-shifted due to the coupling with the 
$K^{\pi}=0^{+}$ component of the ISGQR.  
Then, as the mass number increases from $A=100$ to 112, 
the excitation energy is downward-shifted. 
The difference in the excitation energy of the upper peak in $^{100}$Zr and in $^{112}$Zr 
is 0.7 MeV. 
This energy difference is understood by 
the systematic trend of the excitation energy $E_{x} \propto A^{-1/3}$.

The lower peak of the ISGMR in deformed systems 
originates from the ISGQR in the spherical limit. 
The excitation energy of the lower peak in the scaling model agrees well 
with that obtained in the present calculation. 

Figure~\ref{energy}(b) shows the fraction of the EWSR value 
for the lower and upper peaks of the ISGMR. 
The energy interval to evaluate the energy weighted strength 
is the same as in the evaluation of the mean excitation energy. 
The values obtained in the scaling model~\cite{nis85} are also shown. 
Unlike for the excitation energy, we can clearly see deviation from 
the values obtained in the scaling model as the neutron number increases. 
We have an appreciable enhancement of the transition strengths of the lower peak. 
One may attribute the deviation to the neutron excess.

To clearly see the effect of the neutron excess, we move on to 
the drip-line systems. 
Figure~\ref{response2} shows the IS and IV monopole and quadrupole transition-strength 
distributions in $^{136,138,140}$Zr. 
We can see the further enhancement of the IS monopole transition strength 
in the lower peak region. 
The mean energy of the lower and upper peaks is shown in Fig.~\ref{energy}(a). 
We set $E_{c}$ to 13 MeV. 
Even in the drip-line nuclei, 
we can not see an appreciable deviation from the simple scaling model 
both for the mean energies of the lower and upper peaks.

However, the fraction of the EWSR value of the lower peak 
keeps increasing toward a drip line. 
Note that the deformation of near-drip-line nuclei $^{136,138,140}$Zr 
is smaller than that of $^{108,110,112}$Zr. 
In the scaling model, the coupling between the GMR and GQR is governed by 
the quadrupole deformation of the ground state~\cite{nis85}. 
Thus, the calculated fraction of the EWSR value of the lower peak 
in the scaling model is small in the $A \sim 140$ region.

To investigate the origin of the enhancement of the transition strengths 
in the low energy region, we show 
in the middle panels of Figs.~\ref{response1} and \ref{response2} 
the neutron and proton monopole transition strengths. 
It is seen that the neutron excitation plays a dominant role 
in increasing the IS transition strengths in the energy region around 10 MeV. 

\begin{table}[t]
\begin{center}
\caption{Sum of the transition strengths of neutrons and protons 
$M_{\nu}^{2}$, $M_{\pi}^{2}$ (in $10^{3}$ fm$^{4}$) and 
$M_{\nu}/M_{\pi}$ divided by the neutron and proton numbers 
for the lower and upper peaks of the ISGMR.}
\label{MnMp}
\begin{tabular}{ccccccc}
\hline \hline
\noalign{\smallskip}
  & \multicolumn{3}{c}{Lower peak} & \multicolumn{3}{c}{Upper peak}  \\
  & $M_{\nu}^{2}$ & $M_{\pi}^{2}$ & $\dfrac{M_{\nu}/M_{\pi}}{N/Z}$ 
& $M_{\nu}^{2}$ & $M_{\pi}^{2}$ & $\dfrac{M_{\nu}/M_{\pi}}{N/Z}$ \\
\noalign{\smallskip}\hline\noalign{\smallskip}
$^{90}$Zr &  &  &  & 2.62 & 1.70 & 0.99 \\
$^{100}$Zr & 2.33 & 0.75 & 1.18 & 2.75 & 1.44 & 0.92  \\
$^{108}$Zr & 3.53 & 0.76 & 1.27 & 3.68 & 1.43 & 0.94  \\
$^{110}$Zr & 3.64 & 0.77 & 1.24 & 3.85 & 1.48 & 0.92  \\
$^{112}$Zr & 4.50 & 0.85 & 1.28 & 3.93 & 1.54 & 0.89  \\
$^{136}$Zr & 9.35 & 0.60 & 1.64 & 5.82 & 1.78 & 0.75  \\
$^{138}$Zr & 10.2 & 0.65 & 1.62 & 5.83 & 1.82 & 0.73  \\
$^{140}$Zr & 10.8 & 0.72 & 1.55 & 6.05 & 1.86 & 0.72  \\
\noalign{\smallskip}
\hline \hline
\end{tabular}
\end{center}
\end{table}

Table~\ref{MnMp} summarizes the contribution of the neutrons and protons 
to the lower and upper peaks of the ISGMR. 
We can see the effect of the neutron excess on the GMR quantitatively from this Table. 
Listed in this Table are the sum of the transition strengths of 
neutrons $M_{\nu}^{2}$ and protons $M_{\pi}^{2}$ 
in the energy intervals of $5$ MeV $<\hbar \omega_{i}< E_{c}$ and 
$E_{c}<\hbar \omega_{i}<$ 25 MeV. 
If the neutrons and protons coherently contribute to the excitation mode, 
the ratio of the matrix elements of neutrons to protons may approach 
the ratio of the neutron number to the proton number. 

The ratio of the matrix elements divided by the ratio of neutron to proton numbers, 
$(M_{\nu}/M_{\pi})/(N/Z)$, for the lower peak increases in the drip line nuclei. 
On the other hand, the quantity $(M_{\nu}/M_{\pi})/(N/Z)$ 
for the upper peak decreases in the drip line nuclei. 
Thus, the neutron excitation concentrates particularly in the energy region of 
the lower peak of the ISGMR. 
Besides the GMR, we can see appearance of the threshold strength consisting 
of neutron excitations only in the drip-line nuclei. 
This fact is similar to the findings of Ref.~\cite{ham97} in $^{60}$Ca.

We are going to move on to the discussion on the properties 
of the IVGMR. 
In $^{90}$Zr, the mean energy of the IVGMR is 31.6 MeV. 
The hydrodynamical model taking a surface effect into account~\cite{bow87} predicts 
the excitation energy of the IVGMR to be
\begin{equation}
E_{x}\simeq 88A^{-1/6}(1+\dfrac{14}{3}A^{-1/3})^{-1/2} \hspace{0.1cm} \mathrm{MeV}.
\end{equation}
This gives the energy of the IVGMR in $^{90}$Zr to be about 29 MeV. 
This is consistent with the microscopically calculated energy. 

The IVGMR in deformed neutron-rich nuclei 
has interesting features, where a four-peak structure appears 
as shown in Fig.~\ref{response1}. 
The four-peak structure is clearly seen especially in $^{112}$Zr. 

The highest peak above 30 MeV corresponds to a primal IVGMR 
by comparing the peak energy with that in $^{90}$Zr. 
We have a peak structure at $22-23$ MeV where the IVGQR is seen. 
Thus the second highest peak emerges associated with the coupling 
to the $K^{\pi}=0^{+}$ component of the IVGQR.
 
Besides these two peaks, we can see two more peaks below 20 MeV. 
We can consider that they are brought to the energy region of the ISGMR 
due to the neutron excess. 
One is associated with the coupling to the ISGMR, 
and the other is due to the coupling to the $K^{\pi}=0^{+}$ component 
of the ISGQR. 

In the drip-line nuclei, 
the transition strengths concentrate in the ISGMR region. 
We can still see a four-peak structure in $^{136,138,140}$Zr. 
The IV transition strengths become enhanced in the energy region 
where the neutron transition strengths increase. 
Therefore, we can see the threshold strength in the IV monopole 
strength distribution as well.
 
\section{Summary}\label{summary}
We have investigated the deformation effects on the 
GMR by employing the newly developed parallelized QRPA calculation scheme.
Associated with the deformation, 
the excitation modes with the angular momenta $l=0$ and $l=2$ can mix 
in the $K^{\pi}=0^{+}$ channel, 
and accordingly we have obtained a two-peak structure of the ISGMR. 
The lower peak of the ISGMR is due to the coupling with the ISGQR.
The deformation splitting is seen in nuclei from near-stability line 
to drip line. 

The deformation splitting obtained in the present calculation 
is in fairly good agreement with the result of the scaling model, 
while the distribution of the strengths 
is quite different when the drip line is approached. 
The IS monopole transition strengths are enhanced in the energy region of the 
lower peak of the ISGMR. 
This is due to the concentration of the neutron transition strengths 
in the lower energy region. 

We have found a complicated peak structure for the IVGMR in deformed neutron-rich nuclei, 
where a four-peak structure appears. 
The highest peak corresponds to a primal IVGMR, and the second is associated 
with the coupling to the $K^{\pi}=0^{+}$ component of the IVGQR, and 
the two more peaks are found in the ISGMR region 
associated with the neutron excess. 

Besides the GMRs, the threshold strength is found in the drip-line nuclei. 
The strength consists of neutron strengths only. 
Since the low-lying $K^{\pi}=0^{+}$ mode is quite sensitive to the shell structure 
and surface structure~\cite{yos08b}, we are going to investigate 
the microscopic structure of the low-lying $K^{\pi}=0^{+}$ states 
with special attention to 
the coupling to the continuum states and the pairing correlations. 

\begin{acknowledgments}  
The author thanks K.~Matsuyanagi for stimulating discussions and useful comments. 
He is supported by the Special Postdoctoral Researcher Program of RIKEN. 
The numerical calculations were performed 
on RIKEN Integrated Cluster of Clusters (RICC) and 
on the NEC SX-8 supercomputers
at the Yukawa Institute for Theoretical Physics, Kyoto University and 
at the Research Center for Nuclear Physics, Osaka University.
\end{acknowledgments}


\begin{thebibliography}{99}
\bibitem{paa07}
N.~Paar, D.~Vretenar, E.~Khan and G.~Col\`o,
Rep. Prog. Phys. {\bf 70}, 691 (2007).

\bibitem{yos09b}
K.~Yoshida, Phys. Rev. C {\bf 80}, 044324 (2009).

\bibitem{har01}
M.~N.~Harakeh and A. van der Wounde,
{\it Giant Resonances: Fundamental High-Energy Modes of Nuclear Excitation} 
(Oxford, 2001).

\bibitem{BM2}
A.~Bohr and B.~R.~Motteleson, 
{\it Nuclear Structure}, vol.~II (Benjamin, 1975; World Scientific, 1998).

\bibitem{gar80}
U.~Garg {\it et al}., Phys. Rev. Lett. {\bf 45}, 1670 (1980).

\bibitem{ben03}
M.~Bender, P-H.~Heenen, P-G.~Reinhard, 
Rev. Mod. Phys. 75 (2003) 121.

\bibitem{paa10}
N.~Paar, J. Phys. G: Nucl. Part. Phys. {\bf 37}, 064014 (2010). 

\bibitem{per08}
S.~P\'eru and H.~Goutte, Phys. Rev. C {\bf 77}, 044313 (2008).

\bibitem{nes06}
V.~O.~Nesterenko, W.~Kleinig, J.~Kvasil, P.~Vasely, P.~G.~Reinhard, and D.~S.~Dolci, 
Phys. Rev. C {\bf 74}, 064306 (2006).

\bibitem{bla05}
A.~Blazkiewicz, V.~E.~Oberacker, and A.~S.~Umar, and M.~Stoitsov, 
Phys. Rev. C {\bf 71}, 054321 (2005).

\bibitem{sto08}
M.~Stoitsov, N.~Michel, and K.~Matsuyanagi, Phys. Rev. C {\bf 77}, 054301 (2008).

\bibitem{yos10}
K.~Yoshida and T.~Nakatsukasa, in preparation.

\bibitem{yos08}
K.~Yoshida and N.~V. Giai, Phys. Rev. C {\bf 78}, 064316 (2008).

\bibitem{bul80}
A.~Bulgac, Preprint No. FT-194-1980, 
Institute of Atomic Physics, Bucharest, 1980. 
[arXiv:nucl-th/9907088]

\bibitem{dob84}
J.~Dobaczewski, H.~Flocard and J.~Treiner,
Nucl. Phys. {\bf A422}, 103 (1984).

\bibitem{bar82}
J.~Bartel, P.~Quentin, M.~Brack, C.~Guet, and H.-B.~H\r{a}kansson, 
Nucl. Phys. {\bf A386}, 79 (1982).

\bibitem{cha76}
R.~R.~Chasman, Phys. Rev. C {\bf 14}, 1935 (1976).

\bibitem{ben05}
K.~Bennaceur and J.~Dobaczewski, Comput. Phys. Commun. {\bf 168}, 96 (2005).

\bibitem{row70}
D.~J.~Rowe, {\it Nuclear Collective Motion}, (Methuen and Co. Ltd., 1970).

\bibitem{ter05}
J.~Terasaki, J.~Engel, M.~Bender, J.~Dobaczewski, W.~Nazarewicz, and
M.~Stoitsov, Phys. Rev. C {\bf 71}, 034310 (2005).

\bibitem{eba10}
S.~Ebata, T.~Nakatsukasa, T.~Inakura, K.~Yoshida, Y.~Hashimoto, and K.~Yabana, 
arXiv:1007.0785.

\bibitem{ina09}
T.~Inakura, T.~Nakatsukasa, and K.~Yabana, 
Phys. Rev. C {\bf 80}, 044301 (2009).

\bibitem{nis85}
S.~Nishizaki and K.~And\=o, Prog. Theor. Phys. {\bf 73}, 889 (1985).

\bibitem{ham97}
I.~Hamamoto, H.~Sagawa, and X.~Z.~Zhang, 
Phys, Rev. C {\bf 56}, 3121 (1997).

\bibitem{bow87}
J.~D.~Bowman, E.~Lipparini and S.~Stringari, 
Phys. Lett. {\bf B197}, 497 (1987).

\bibitem{yos08b}
K.~Yoshida and M.~Yamagami, Phys. Rev. C {\bf 77}, 044312 (2008).

\end{thebibliography}
\end{document}